\newcommand{\foiii}{[O\thinspace{\sc iii}]}

\newcommand{\foii}{[O\thinspace{\sc ii}]}
\newcommand{\fsii}{[S\thinspace{\sc ii}]}
\newcommand{\fsiii}{[S\thinspace{\sc iii}]}

\newcommand{\fnii}{[N\thinspace{\sc ii}]}

\newcommand{\fcliii}{[Cl\thinspace{\sc iii}]}

\newcommand{\oii}{O\thinspace{\sc ii}}

\newcommand{\cii}{C\thinspace{\sc ii}}

\documentclass[twoside,a4paper,11pt]{sea10}
\usepackage{graphicx}
\usepackage{hyperref}
\usepackage{movie15}
\begin{document}
\pagenumbering{arabic}
\pagestyle{myheadings}
\thispagestyle{empty}
{\flushleft\includegraphics[width=\textwidth,bb=58 650 590 680]{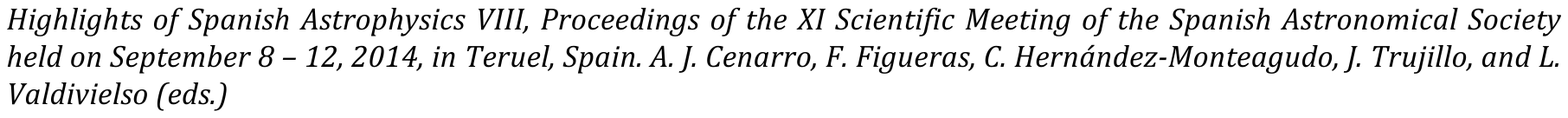}}
\vspace*{0.2cm}
\begin{flushleft}
{\bf {\LARGE
%
Carbon abundances and radial gradients in NGC\,300 and other nearby spiral galaxies
%
}\\
\vspace*{1cm}
%
L. Toribio San Cipriano$^{1,2}$,
J. Garc\'ia-Rojas$^{1,2}$, 
and 
C. Esteban$^{1,2}$
%
}\\
\vspace*{0.5cm}
%
$^{1}$
Instituto de Astrof\'isica de Canarias, E- 38200 La Laguna, Tenerife, Spain\\
$^{2}$
Departamento de Astrof\'isica, Universidad de La Laguna, E-38071 La Laguna, Tenerife, Spain\\

%
\end{flushleft}
%
\markboth{
Carbon abundances and radial gradients
}{ 
%
L. Toribio San Cipriano et al. 
%
}
\thispagestyle{empty}
\vspace*{0.4cm}
\begin{minipage}[l]{0.09\textwidth}
\ 
\end{minipage}
\begin{minipage}[r]{0.9\textwidth}
\vspace{1cm}
\section*{Abstract}{\small
%

We present preliminary results of deep echelle spectrophotometry of a sample of HII regions along the disk of the Scd galaxy NGC\,300 obtained with the Ultraviolet and Visual Echelle Spectrograph (UVES) at the Very Large Telescope (VLT) with the aim of detect and measure very faint \oii\ and \cii\ permitted lines. We focus this study on the C and O abundances obtained from faint optical  recombination lines (ORLs) instead of the most commonly used collisionally excited lines (CELs). We have derived the ionic abundances of $\mathrm{C^{2+}}$ from the \cii\ 4267$\mathrm{\AA}$ RL and $\mathrm{O^{2+}}$ from the multiplet 1 of \oii\ around 4649$\mathrm{\AA}$ in several objects. 
Finally, we have computed the radial gradients of C/H, O/H and C/O ratios  in NGC\,300 from RLs, which has allowed the comparison with similar data obtained by our group in other nearby spiral galaxies.  
%
\normalsize}
\end{minipage}
%
%
%
\section{Introduction \label{intro}}

The study of chemical abundances and their distribution across the disk of galaxies provides essential information to understand the nuclear processes in stellar interiors and the chemical evolution and formation of galaxies. The main aim of this work is to determine and analyse $\mathrm{C/H}$, $\mathrm{O/H}$ and $\mathrm{C/O}$ radial abundances gradients using optical recombinations lines (ORLs) from deep HII regions spectra in nearby galaxies.

We pay special attention to carbon (C) because it is the second most abundant heavy-element in the Universe and is of paramount biogenic importance.  
Unfortunately, despite its importance, there are few studies about C abundances in extragalactic HII regions and mainly derived from UV collisionally excited lines, which are extremely sensitive to precise determinations of extinction and electron temperature of the nebula.

\section{The data}

We selected seven HII regions of NGC\,300 from the samples of \cite{Bresolin09} and \cite{Deharveng88}, according to their high surface brightness and position along the disk of the galaxy.  These regions were observed at Cerro Paranal Observatory using Ultraviolet Visual Echelle Spectrograph (UVES) at the VLT in service time between 2010 July and August. We covered the whole optical spectral range and integrated 1.75 h on each region, with a slit width of 3$''$. In this work we also present additional deep spectrophotometric data of HII regions from previous studies by our group of the Milky Way and other nearby spiral galaxies: M\,31, M\,101, M\,33 and NGC\,2403. These data were taken with some of the largest aperture telescopes available: VLT, KECK and SUBARU telescopes.

\section{Line  intensities}

To measure the line fluxes we used the SPLOT routine of IRAF. We  integrated all the flux in the line between two given limits and over a local continuum estimated by eye.  In case of line blending, we fitted a double Gaussian profile to measure the individual line intensities.

All line fluxes were normalized to $\mathrm{H}_\beta$ and corrected for interstellar reddening. The identification and laboratory wavelength of the lines were obtained following the previous works of \cite{Garcia-Rojas04} and \cite{Esteban04}

\begin{figure}[h]
\center
\includegraphics[scale=0.4]{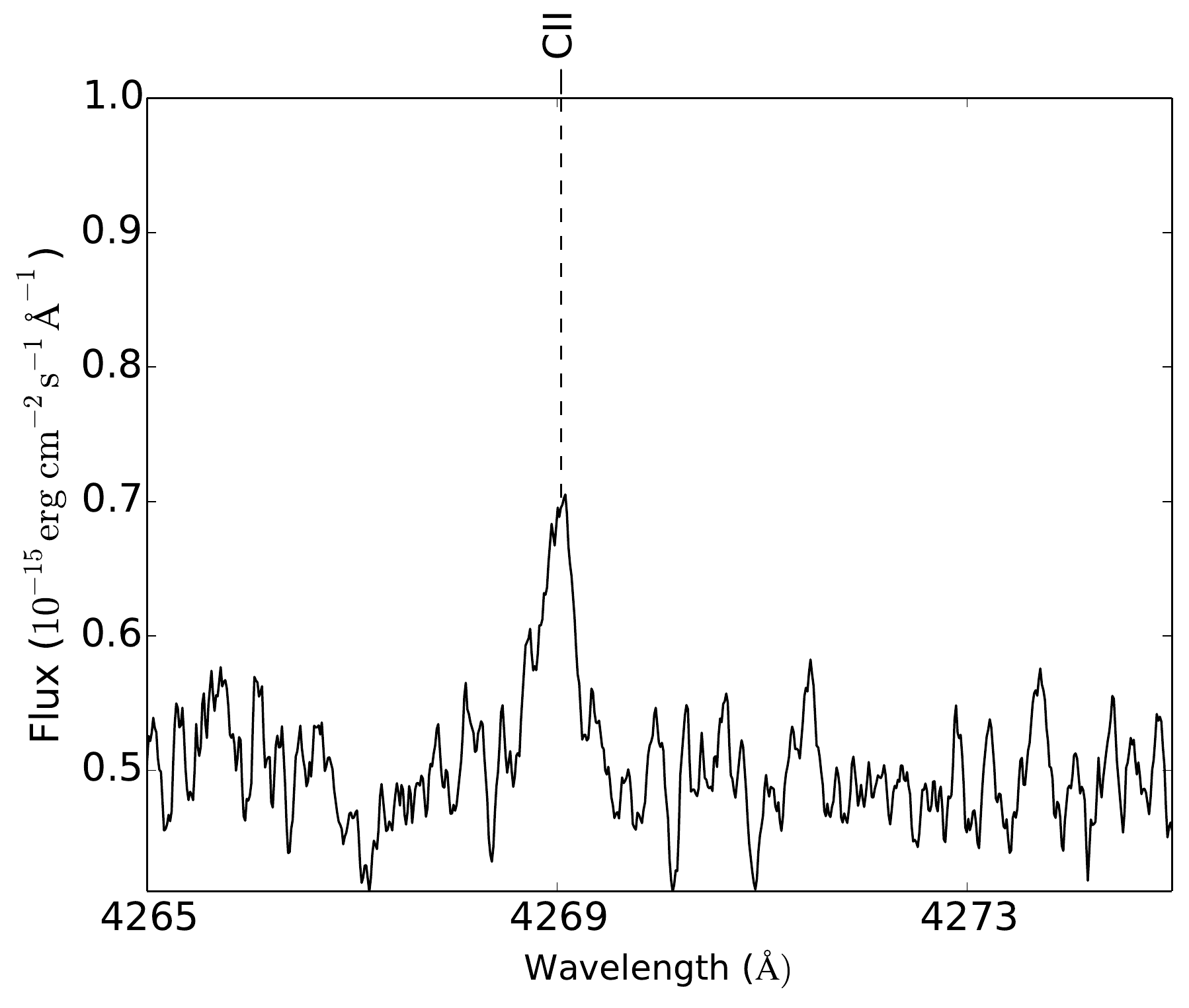} 
\includegraphics[scale=0.4]{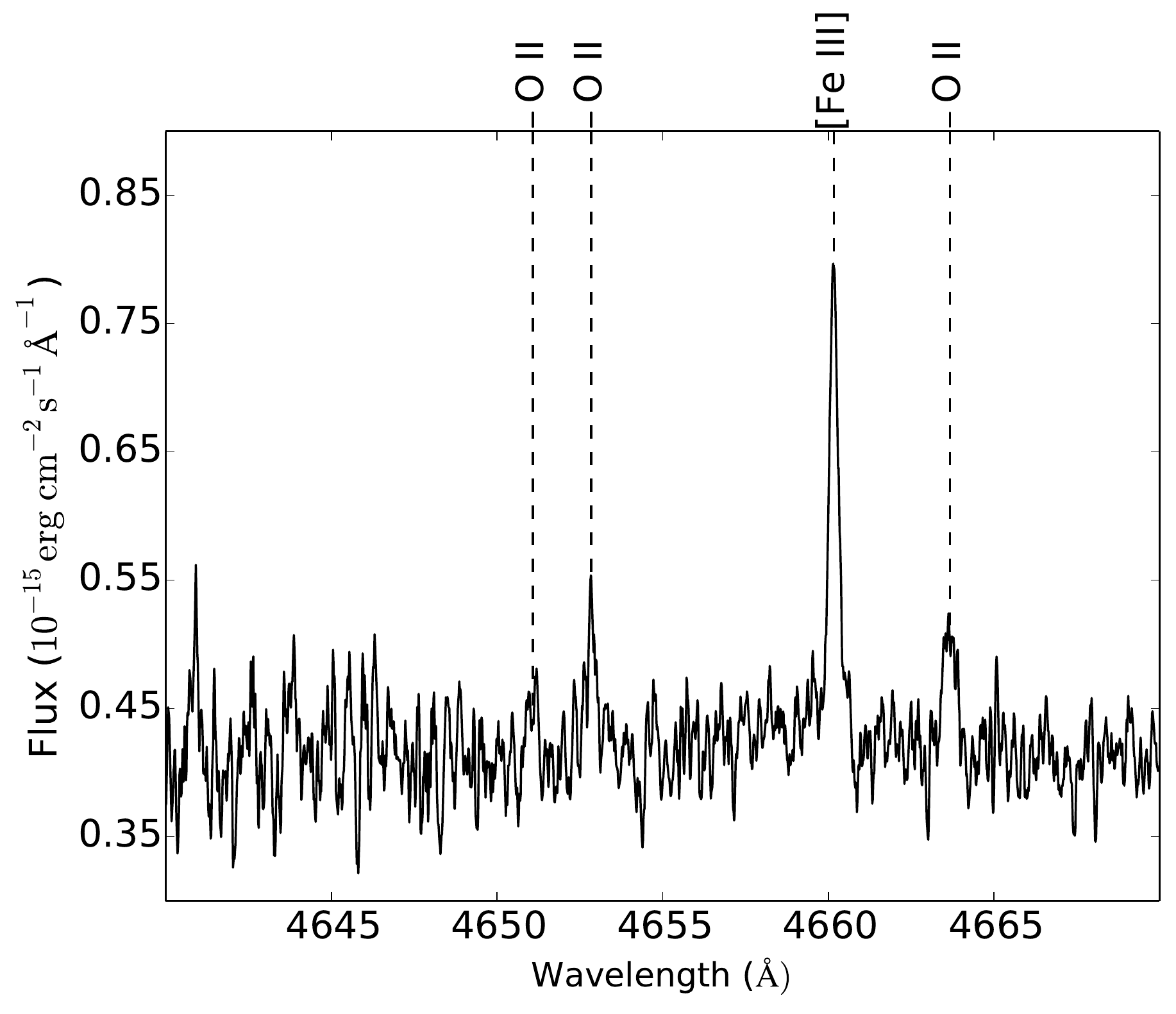} 
\caption{\label{fig1} Sections of the spectra of the HII region R14 in NGC\,300. The panels show \cii\ $\lambda$4267 (left) recombination line and the lines of multiplet 1 of \oii\ $\lambda\lambda$4650 (right).}
\end{figure}

The \cii\ $\lambda$4267 line and some lines of multiplet 1 of \oii\ around $\lambda$4649  were detected in 4 regions. Fig. \ref{fig1} show the spectral ranges covering \cii\  $\lambda$4267 (left panel) and multiplet 1 of \oii\ (right panel) of the HII region R14.

\section{Physical conditions}

The physical conditions of the ionized gas, electron density and temperature,  were derived using PyNeb, a Python-based and updated expansion of the \thinspace{\sc{iraf}} package \thinspace{\sc{nebular}} (\cite{Luridiana14}).

Electron density, $n_\mathrm{e}$ was derived from \fsii\ 6731/6716, \foii\ 3726/3729 and \fcliii\ 5538/5518 lines ratios.  Electron temperature, $T_\mathrm{e}$ was derived from  \foiii\ 4363/5007, \fsiii\ 6312/9069, \fnii\ 5755/6584, \foii\ (7320+30)/(3726+29) and \fsii\ (4068+76)/(6717+31) lines ratios. We assumed a two-zone scheme, where the high ionization zone is characterized by a $T_\mathrm{e}$ that is the  weighted mean of T$_\mathrm{e}$(\foiii) and T$_\mathrm{e}$(\fsiii) whereas for the low-ionization zone, we computed $T_\mathrm{e}$ as the weighted mean of T$_\mathrm{e}$(\fnii) and T$_\mathrm{e}$(\foii). The uncertainties on the temperatures were computed through Monte Carlo simulations.

\section{Preliminary results}

\subsection{Chemical abundances}

\begin{figure}[]
\center
\includegraphics[scale=0.42]{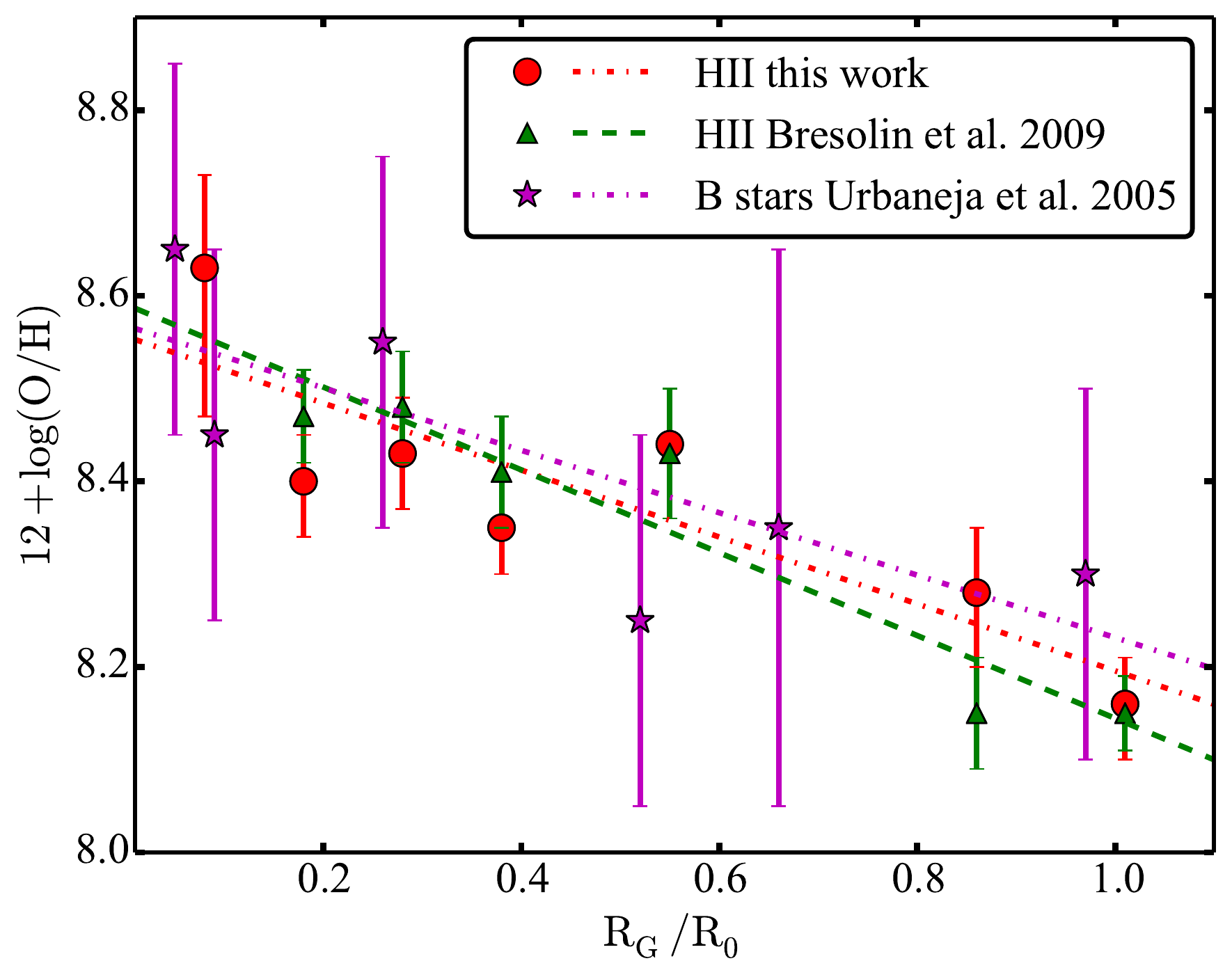} 
\includegraphics[scale=0.42]{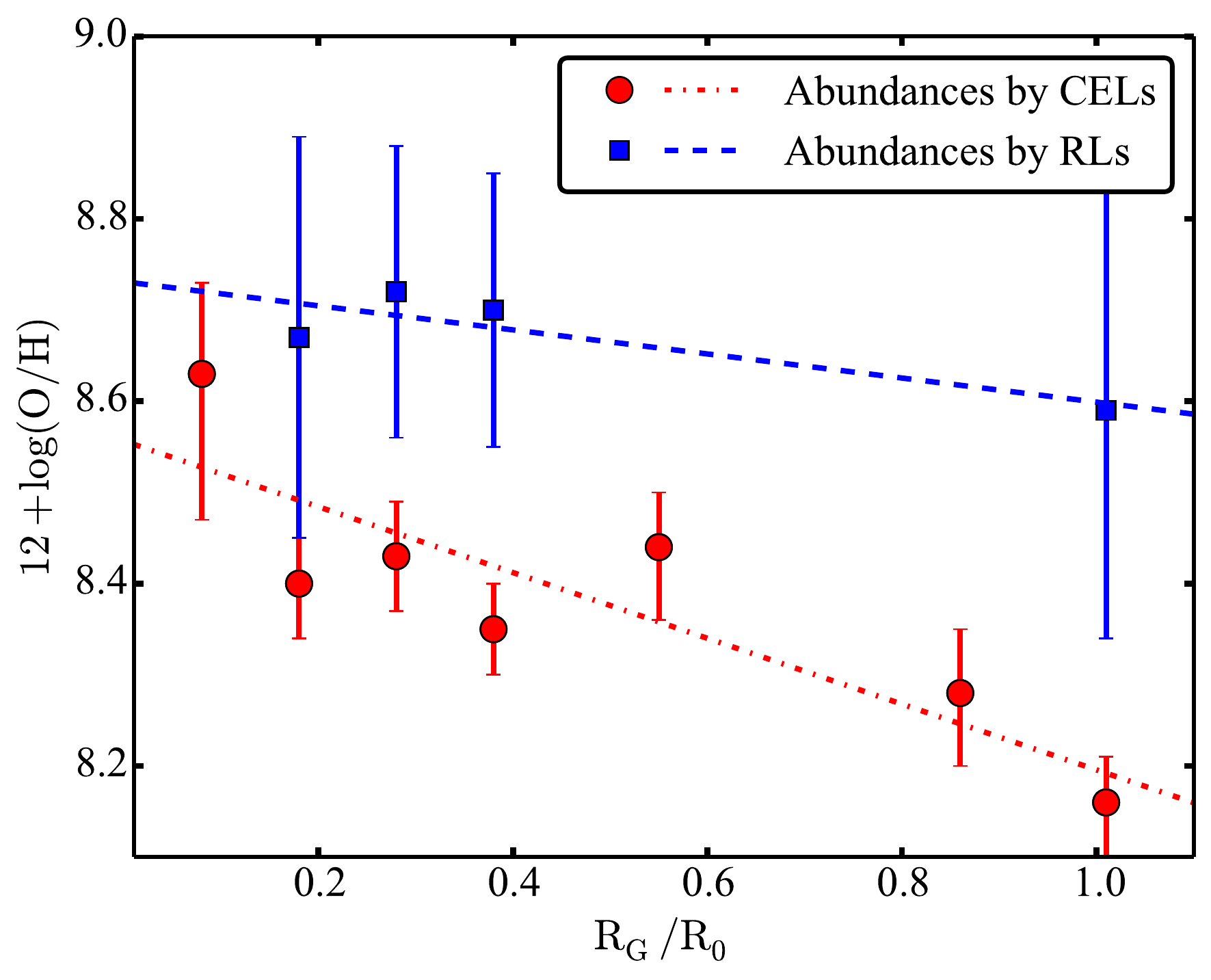} 

\caption{\label{fig2} Left panel shows the comparison of radial $\mathrm{O/H}$ gradients obtained from our HII regions data (red circles), HII regions data by \cite{Bresolin09} (green triangles) and B stars data by \cite{Urbaneja05} (purple stars). Right panel shows the comparison between the radial O/H gradient we obtain from  CELs (red circles) or from ORLs (blue squares).}
\end{figure}

Traditionally, the standard method for deriving ionic abundances is based in the intensity of bright CELs. This lines are much brighter than ORLs, and therefore easier to detect. However, abundances relative to hydrogen derived from CELs are strongly affected by uncertainties in the electron temperature. On the other hand, abundances derived from ORLs are almost independent on the physical conditions, but in extragalactic HII regions these lines are extremely difficult to observe.

We derived $\mathrm{O/H}$ ratios for each HII region using both CELs and ORLs (when available). $\mathrm{C/H}$ ratios can only be obtained from ORLs and, for consistency $\mathrm{C/O}$ ratios were computed also from ORLs. To compute total O/H ratio from CELs, we simply added O$^+$ and O$^{++}$ abundances. For ORLs, where we could not compute O$^+$ from these kind of lines, we assumed the $\mathrm{O^+/O^{++}}$ ratio determined from CELs  to obtain the final O/H ratio. To take into account for the unseen ionization stages of C, we used the ionization correction factor (ICF) proposed by \cite{Garnett99} to compute the total C abundance. In the left panel of Fig.~\ref{fig2}, we compare the radial O/H gradient we derive from CELs with the radial O/H gradients obtained by \cite{Bresolin09} from HII regions and by \cite{Urbaneja05} from the analysis of B stars. It can be seen than our results are consistent, within the errors, with those derived by these authors. In the right panel of Fig. \ref{fig2} we compare the galactic radial $\mathrm{O/H}$ gradients obtained using CELs and ORLs. The O abundances based on ORLs are systematically higher than those derived from CELs. This is the so-called {\it Abundance Discrepancy problem}. Today, this is one of the key problems in the physics of photoionized nebulae.

\subsubsection{Correlations between O/H, C/H and C/O radial gradients with some properties of the galaxies}

\begin{figure}
\center
\includegraphics[scale=0.4]{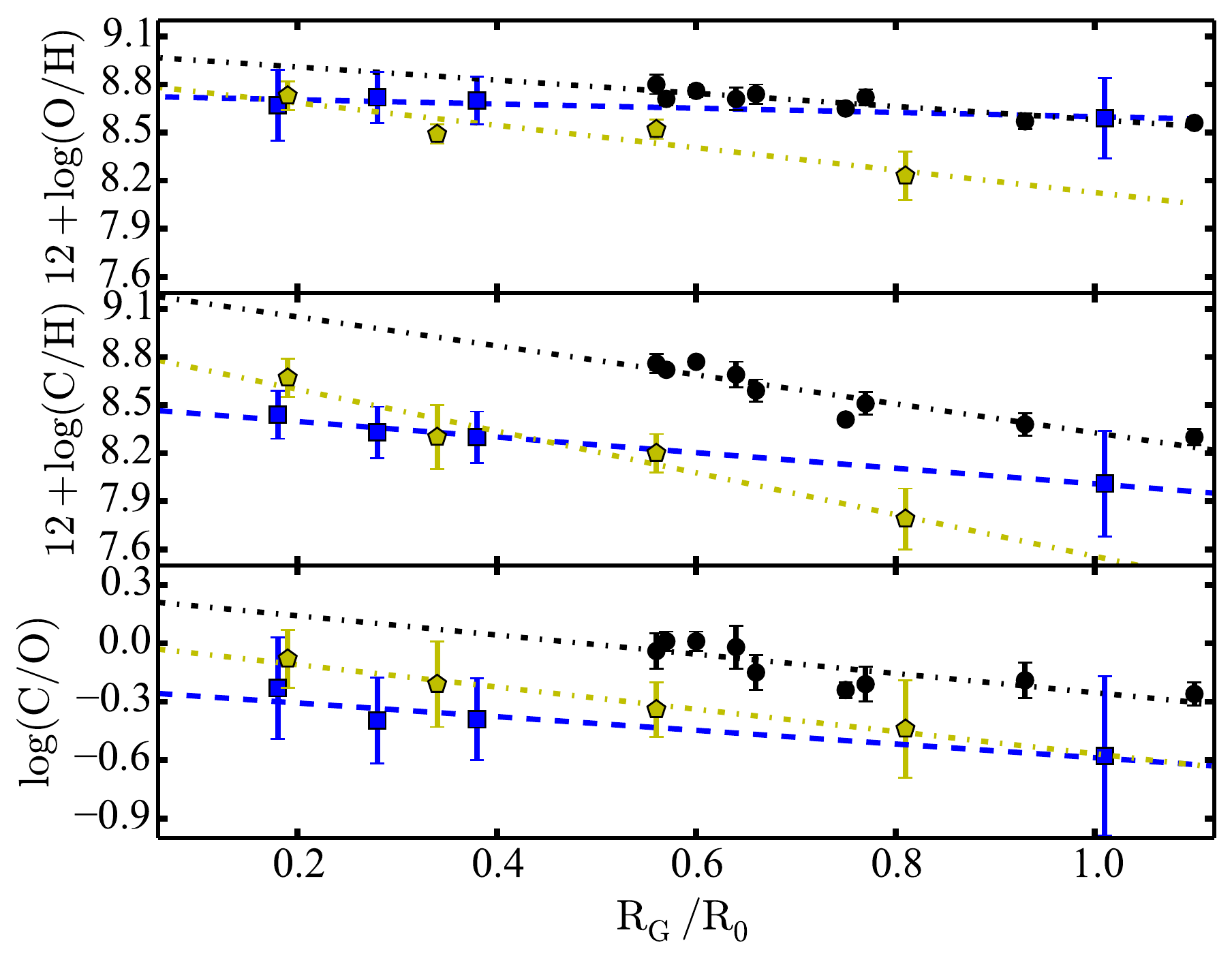} 
\includegraphics[scale=0.4]{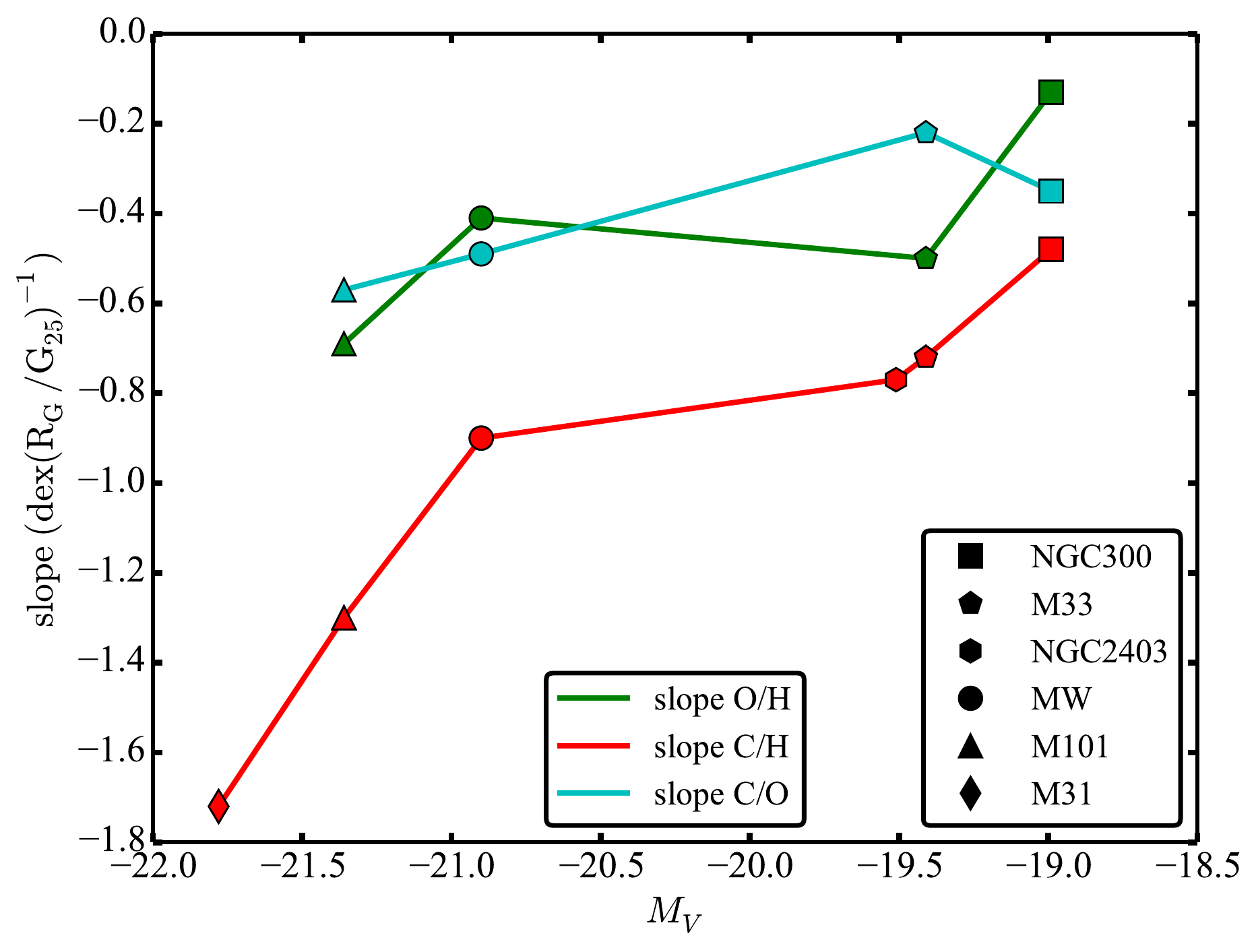} 
\caption{\label{fig3} Lelf panel shows the  $\mathrm{O/H}$ ratio (upper), $\mathrm{C/H}$ ratio (center) and $\mathrm{C/O}$ ratio (lower) radial gradients for NGC\,300 (blue squares), Milky Way (black circles) and M\,101 (yellow pentagons). In the right panel, we show the slope of  $\mathrm{O/H}$ (green), $\mathrm{C/H}$ (red) and $\mathrm{C/O}$ (blue) radial gradients $vs.$ the absolute magnitude. $M_V$, for NGC\,300 (squares), M\,33 (pentagons), NGC\,3403 (hexagons), Milky Way (circles), M\,101 (triangles) and M\,31 (diamonds).}
\end{figure}

In Table~\ref{tab1} we summarize the results obtained for NGC\,300 as well as previous results obtained by our group for other nearby spiral galaxies. In the left panel of Fig.~\ref{fig3}, we show the radial abundance gradients of $\mathrm{O/H}$, $\mathrm{C/H}$ and $\mathrm{C/O}$ we derive for NGC\,300 from ORLs together with the results obtained by \cite{Esteban05, Esteban09, Esteban13} for the Milky Way and the spiral galaxy M\,101. From this figure and Table~\ref{tab1}, we can see that the C/H gradients are steeper that the $\mathrm{O/H}$ ones in all galaxies, leading to negative gradients of the $\mathrm{C/O}$ ratio. Moreover, there seems to be a correlation between steeper gradients and the luminosity (mass) of the spiral galaxy. The correlation is more evident in the $\mathrm{C/H}$ and $\mathrm{C/O}$ gradients. This result can be clearly seen in right panel of Fig.~\ref{fig3} where the slope of  $\mathrm{O/H}$, $\mathrm{C/H}$ and $\mathrm{C/O}$ radial gradients for different nearby spiral galaxies is represented versus their absolute magnitude, $M_V$, which is an indicator of the luminosity of the galaxy. The correlation is more evident in the $\mathrm{C/H}$  gradient. This is an important result that we will try to confirm with further observations of HII regions in M101, M33 and other nearby spiral galaxies.

\begin{table}[ht] 
\caption{Summary of properties of the galaxies shown in Fig.~\ref{fig3}.} 
\center
\begin{tabular}{cccccc} 
\hline\hline 
           & Morphological& &\multicolumn{3}{c}{\underline{slope($\mathrm{dex(R_G/R_{25})^{-1}}$)}}\\
Galaxy &  type  & $\mathrm{M_v}$ & $\mathrm{O/H}$&$\mathrm{C/H}$&$\mathrm{C/O}$\\
\hline 
NGC300     & Scd  & -18.99 & -0.13 & -0.48 & -0.35 \\
M33           & Sc    & -19.41 & -0.50 & -0.72 & -0.22\\
NGC2403   & Sc   & -19.51 & --       & -0.77 & -- \\
MW            & SBc  & -20.90 & -0.41 & -0.90  & -0.49 \\
M101         & SBc  &-21.36 & -0.69 & -1.30 & -0.57 \\
M31           & SAb  &-21.78 & --         & -1.72 & --\\
\hline
\end{tabular} 
\label{tab1} 
\end{table}
%
%
\small  
%
\section*{Acknowledgments}   
%
This work is based on observations collected at the European Southern Observatory, Chile. We acknowledge funding from the Spanish Ministerio de Econom\'ia y Competividad (MINECO) under grant AYA2011-22614.
%

%
\end{document}